\documentclass[showpacs,preprintnumbers,amsmath,amssymb,twocolumn,prb]{revtex4}
\usepackage{graphicx}% Include figure files
\usepackage{dcolumn}% Align table columns on decimal point
\usepackage{bm}% bold math
\usepackage{bbm}
\usepackage{amsmath}

\begin{document}

\title{Self-energy feedback and frequency-dependent interactions in the functional renormalization group flow for the two-dimensional Hubbard model}% Force line breaks with \\

\author{Stefan Uebelacker}
\email{uebelacker@physik.rwth-aachen.de}
\author{Carsten Honerkamp}
\affiliation{%
Institute for Theoretical Solid State Physics, RWTH Aachen University, D-52056 Aachen 
\\ and JARA - FIT Fundamentals of Future Information Technology
}%

\date{September 18, 2012}% It is always \today, today,

\begin{abstract}
We study the impact of including the self-energy feedback and frequency-dependent interactions on functional renormalization group grows for the two-dimensional Hubbard model on the square lattice at weak to moderate coupling strength. Previous studies using the  functional renormalization group had ignored these two ingredients to large extent, and the question is how much the flows to strong coupling analyzed by this method depend on these approximations. Here we include the imaginary part of the self-energy on the imaginary axis and the frequency-dependence of the running interactions on a frequency mesh of 10 frequencies on the Matsubara axis. We find that i) the critical scales for the flows to strong coupling are shifted downwards by a factor that is usually of order one but can get larger in specific parameter regions, and ii) that the leading channel in this flow does not depend strongly on whether self-energies and frequency-dependence is included or not. We also discuss the main features of the self-energies developing during the flows.
\end{abstract}


\maketitle

\section{\label{intro}Introduction}
Weakly to moderately correlated, itinerant electron systems like iron arsenide superconductors or carbon materials are in the center of todays condensed matter research. Nevertheless, a detailed quantitative description of their low-temperature phases is often not available. Two potential obstacles are that these materials, despite not being strongly correlated, can still host various competing trends that invalidate simpler theoretical approaches, and that the underlying lattice causes anisotropic behavior that can lead to significant quantitative effects such as anisotropic quasiparticle properties or anisotropic energy gaps in the ordered states.
The functional renormalization group (fRG) method has become a widely used tool for the investigation of weakly coupled fermions on low-dimensional lattices (for a recent review, see \cite{metznerRMP}) as it helps in both these aspects. In contrast to often employed random-phase approximation (RPA) or mean field approaches its is capable of describing competing interactions, and it is also capable of resolving stronger momentum-dependencies of the basic obervables like pairing gaps etc.  While fRG applications in two dimensions mainly focussed on the Hubbard model on the square lattice\cite{zanchi,rg1,rg2,rg3}, more recent studies analyze also multiband models, in particular for iron arsenide superconductors\cite{wang,platt,thomale} and graphene systems\cite{honhon,kiesel,scherer}. The bulk of these fRG works uses an approximation to the full flow equations in one which neglects the self-energy completely. The main object is interest in the effective, flowing interaction vertex, whose strongest components give information on the leading correlations in the effective theory at lower scales. Only in a few works, either the flow of the self-energy was computed without allowing for a feedback on the flow of the interaction vertex\cite{kataninkampf,rohe}, or the self-energy was parameterized using a quasiparticle weight\cite{zanchiZ,zflow,katanin2loop}. These studies gave the following information: {\em a)} The fRG flows with neglected self-energy are good approximations in the sense that in typical cases the leading low-energy instabilities are not altered in their existence and character by including (parts of) the self-energy. {\em b)} A detailed study of the self-energy suggests the occurrence of interesting and observable effects such as anisotropic quasiparticle degradation, dispersion renormalizations and partial gap openings. 
Another approximation that has been used in most applications of the fRG is to neglect the frequency-dependence of the interaction vertex. This dependence is usually absent in the initial condition of the flow, i.e. the bare electron-electron interaction is not retarded. But during the flow, the subsequent inclusion induces a frequency-dependence that can cause at least quantitative effects. Again, the frequency-dependence of the vertex has been taken along in a handful of studies\cite{klironomos,fu,honphon}. The upshot here is again that the previous flows without frequency-dependence are not so bad and not drastic changes of the leading instabilities are expected when the approximation is improved. 
  
All these rather qualitative results are encouraging, but in order to develop the fRG into a quantitative method that is able to describe materials more directly, the approach has to be improved. One group of workers have embarked on this mission using ansatzes for the frequency-dependences of vertex functions and self energies\cite{husemann,giering}. Here we want to use a more direct approach. We treat the frequency-dependence of the self-energy and the interaction vertex by discretizing the Matsubara axis in $N_\omega=10$  patches, in a similar spirit as the wavevector-dependence is patched into $N_k=32$ patches around the Fermi surface. This way, the interaction vertex of the one-band model becomes an object with $(N_k \times N_\omega)^3$ components that can still be dealt with numerically. Also, the imaginary part of the self-energy is computed with some resolution, given by $N_\omega$ values on the Matsubara axis. This allows us, e.g., to test how adequate it is to use a simple $Z$-factor in order to describe the flowing self-energy. The self-energy is then fed back into the flow of the interaction vertex by using full Green's functions on the internal lines of the one-loop diagrams in their flow. 

The main questions we can ask using this refined scheme are as follows: {\em a)} What is the impact of the self-energy feedback and/or the frequency-dependence on the type and energy scale of the leading instabilities? Can we use the critical scales in this approximation as estimates for gap sizes in the ordered state at lower scales ? {\em b) } What can be learned about the flowing self-energy? How is its frequency and wavevector-dependence for typical situations in the one-band Hubbard model? Is there any non-Fermi liquid physics that can be deduced from these flows?

\section{\label{method}Model and method}
Throughout this paper we study the two-dimensional Hubbard model on a square lattice with nearest- and next-nearest-neighbor hoppings $t$ and $t^{\prime}$, respectively, and a simple onsite interaction $U$. The Hamiltonian is given by
\begin{equation}
H=\sum_{\mathbf{k},\sigma} \epsilon(\mathbf{k}) c^{\dagger}_{\sigma}(\mathbf{k}) c_{\sigma}(\mathbf{k})+ \sum_{i} U n_{i,\uparrow} n_{i,\downarrow} \; .
\end{equation}
Here $c^{\dagger}_{\sigma}(\mathbf{k})$ and $c_{\sigma}(\mathbf{k})$ are creation and annihilation operators with momentum $\mathbf{k}$ and spin $\sigma \in\{\uparrow, \downarrow\}$ and $n_{i,\sigma}= c^{\dagger}_{i,\sigma} c_{i,\sigma}$ is the density operator at site $i$. The dispersion relation reads as
\begin{equation}
\epsilon(\mathbf{k})=-2t (\cos k_x + \cos k_y) +4 t^{\prime} \cos k_x \cos k_y -\mu \; , 
\end{equation}
where $\mu$ is the chemical potential and the lattice constant is set to unity.

We employ the fRG flow equations for the one-particle-irreducible (1PI) vertex function, which are well explained in detail in the literature (see, e.g., Refs. \onlinecite{salmhoferhonerkamp,metznerRMP,kopietzbook}). The aim is to obtain an effective theory for the low energy degrees of freedom of the system at hand. To do this, one introduces a regulator or cutoff  function $\theta^{\Lambda}(k)$ in the quadratic part of the fermionic action which depends on the RG scale $\Lambda$ so that the free Green's function becomes    
\begin{equation}
G_0^{\Lambda}=\frac{\theta^{\Lambda}(\mathbf{k})}{i \omega-\epsilon(\mathbf{k})}\; ,
\end{equation}
with the Matsubara frequency $\omega$. For the cutoff function $\theta^{\Lambda}(\mathbf{k})$ we choose here a momentum-shell cutoff, 
\begin{equation}
 \theta^{\Lambda}(\mathbf{k})=\Theta(|\epsilon(\mathbf{k})|-\Lambda)\; ,
\end{equation} 
where $\Theta$ is the step function. In the numerical implementation, the step function is slightly softened for better feasibility. The cutoff function suppresses all modes with $|\epsilon(\mathbf{k})|$ below the scale $\Lambda$. At the beginning of the fRG flow at $\Lambda=\Lambda_0 \ge \max(|\epsilon(\mathbf{k})|)$, i.e. greater than the bandwidth, the vertex function of the theory take their bare values, as all perturbative corrections suppressed by the cutoff function. During the fRG flow the scale $\Lambda$ is lowered step by step and in principal we recover the complete action with all vertex corrections at $\Lambda=0$. 

With the so-introduced dependence of the free Green's function on the parameter $\Lambda$ one can derive a differential equation for the generating functional of the 1PI vertices \cite{salmhoferhonerkamp}. After an expansion in the Grassmann fields one arrives at an integro-differential equation for all 1PI vertex functions. In practice we truncate the infinite hierarchy of flow equations by setting the three-particle (six-point) vertex and all higher vertices to zero. In the bare action at $\Lambda=0$ the three-particle vertex vanishes exactly. In principle, during the flow it attains a finite value, which is however neglected in our approximation. Various fRG studies on the Hubbard-model have shown that one can still expect reasonable results as long as one restricts the study to weak or moderate coupling. If the temperature is low enough, a certain part coupling function will typically develop very large values compared to the bandwidth before all modes are integrated out, i.e. will undergo a flow to strong coupling at a nonzero critical scale $\Lambda_c$. Here we have to stop the flow as neglecting the higher-order vertices is no longer justified.  The critical scale $\Lambda_c$ can be used as estimate for ordering temperatures and concomitant spectral modifications, e.g. through gap openings. 

With this approximation and in presence of spin rotational symmetry the fRG differential equation amounts to the following equation for the self- energy $\Sigma^\Lambda$ and two- particle vertex $V^\Lambda$,

\begin{align}
\label{dgl_sigma}
&\frac{d}{d\Lambda}\Sigma^\Lambda \left( k \right) =\int dk' S^\Lambda \left( k' \right)  \Bigl[ V^\Lambda ( k,k',k' ) -2 V^\Lambda ( k,k',k ) \Bigr] \; ,\\
\label{dgl_v}
&\frac{d}{d\Lambda}V^\Lambda ( k_1,k_2,k_3) =\tau^\Lambda_{PP}+\tau^\Lambda_{PH,d}+ \tau^\Lambda_{PH,cr} \; ,
\end{align}
with the particle-particle channel
\begin{align}
\label{pp}
\tau^\Lambda_{PP}&( k_1,k_2,k_3)= \notag\\&- \int dk V^\Lambda ( k_1,k_2,k)L^\Lambda ( k,q_{PP} ) V^\Lambda ( k,q_{PP},k_3) \, , 
\end{align}
the direct particle-hole channel
\begin{align}
\tau&^\Lambda_{PH,d}( k_1,k_2,k_3) = - \int dk \Bigl[\notag\\ &-2V^\Lambda(k_1,k,k_3) L^\Lambda(k,q_{PH,d} ) V^\Lambda(q_{PH,d},k_2,k) \notag\\
&+ V^\Lambda(k,k_1,k_3) L^\Lambda(k,q_{PH,d} ) V^\Lambda(q_{PH,d},k_2,k) \notag\\
&+ V^\Lambda(k_1,k,k_3) L^\Lambda(k,q_{PH,d} ) V^\Lambda(k_2,q_{PH,d},k)    \Bigr] \, , \label{phd} 
\end{align}
and the crossed particle-hole channel
\begin{align}
\label{phcr}
\tau^\Lambda&_{PH,cr}( k_1,k_2,k_3)= \\ &-  \int dk V^\Lambda(k,k_2,k_3) L^\Lambda(k,q_{PH,cr} ) V^\Lambda(k_1,q_{PH,cr},k) \; ,\notag 
\end{align}   
with the combined index $k=(\mathbf{k},\omega)$. Here, $q_{PP}=(-\mathbf{k}+\mathbf{k}_1+\mathbf{k}_2,-w+w_1+w_2)$, $q_{Ph,d}=(\mathbf{k}+\mathbf{k}_1-\mathbf{k}_3;w+w_1-w_3)$, $q_{Ph,cr}=(\mathbf{k}+\mathbf{k}_2-\mathbf{k}_3, w+w_2-w_3)$ are the quantum numbers of the second loop line. We used the shorthand notation for the momentum integral and Matsubara summation $\int dk=\frac{T}{(2 \pi)^2} \int d \mathbf{k} \sum_{\omega}$, with the temperature $T$. The fourth frequency and momentum of the interaction vertex as well as of the second internal line are fixed by conservation. The spin convention is such that the spin is the same for the first and third leg. The internal loop is given by 
\begin{equation}
L^\Lambda(k,k')=S^\Lambda(k)G^\Lambda(k')+G^\Lambda(k)S^\Lambda(k'), \label{loops}
\end{equation}
with the full propagator $G^\Lambda(k)=[(G_0^\Lambda(k))^{-1}-\Sigma^\Lambda(k)]^{-1}$ and the single-scale propagator
\begin{align} \label{singlescale}
S^\Lambda(k)&= -G^\Lambda(k) \left( \frac{d}{d\Lambda}[G_0^\Lambda(k)]^{-1} \right) G^\Lambda(k) \notag \\
&=\frac{[i \omega-\epsilon(\mathbf{k})] \partial_\Lambda \theta^{\Lambda}(\mathbf{k})}{[i\omega-\epsilon(\mathbf{k}) - \theta^{\Lambda}(\mathbf{k}) \Sigma^\Lambda(k)]^2} \; .
\end{align}
Note that we do not employ the Katanin-modification\cite{katanin}, i.e. the replacement of $S^\Lambda(k)$ by $\frac{d}{d\Lambda} G^\Lambda(k)$ in Eq. (\ref{loops}), as this would increase the numerical effort very strongly in this two-dimensional problem.

At the beginning of the flow the coupling function assumes the value of the onsite interaction $V^\Lambda (k_1,k_2,k_3)=U$ and the self-energy vanishes $\Sigma^\Lambda(k)=0$.

\begin{figure}
\begin{center}
\hspace{-1cm}\includegraphics[scale=0.5]{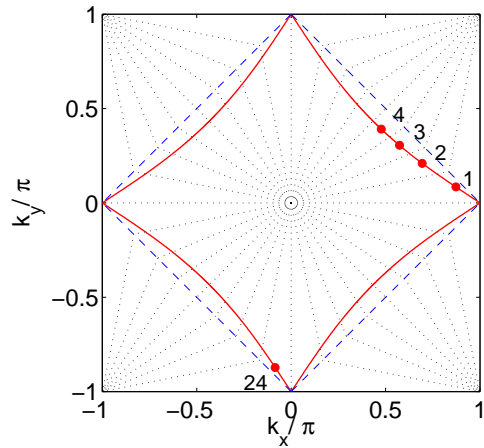}
\end{center}
\caption{N-patch discretization scheme with $N_k=32$. The Fermi surface with $t'=0.2t$, $\mu=-4t'$ is plotted in red. The patches are divided by black dotted lines. The red dots denote the momenta at which the coupling function and self-energy is computed. The umklapp-surface is shown as the blue dashed line.} 
\label{patchingplot} 
\end{figure}

Here we want to study the fRG flow with frequency-dependent self-energy and two- particle vertex. To do this, we divide the Brillouin zone into $N_k$ patches to cover the momentum-dependence of the vertices at the Fermi surface according to Ref. \onlinecite{rg1} (see Fig. \ref{patchingplot}). This amounts to a projection of the momentum-dependence onto the Fermi surface and the vertex has then three momentum indices $\mathbf{k}_i=1,\dots N_k$. Additionally, we divide the Matsubara axis in $N_{\omega}$ patches and calculate the Matsubara sum numerically. In this work, we use two different patching schemes.
In the first scheme (D1) the vertices and self-energy are calculated at the 10 Matsubara frequencies with the smallest absolute values. The Matsubara frequencies of the respective patches are thus given by $\omega_p=\{ \pm 1,  \pm 3, \pm 5, \pm 7, \pm 9 \} \times \pi T$. We keep the smallest frequencies as theses are expected to give the leading contributions to the flow. Additionally we consider a patching where the frequencies of the patches are at $\omega_p=\{ \pm 1,  \pm 3, \pm 5, \pm 11, \pm 21 \} \times \pi T$ to cover the dependence at higher frequencies more accurately. This discretization is referred to as D2. In both schemes the vertices and self-energy at all other frequencies occurring in the loops are assumed to have the same value as the closest frequency-patch. This amounts to keeping the frequency-dependence in self-energy and interactions constant above the largest positive or smallest negative discretization frequency.  

In the loop diagrams with self-energy inclusion, the Matsubara sums have to be perfumed numerically. To do this, we truncate the Matsubara sums at some $\omega_{\infty}$ and check whether further increasing this frequency cutoff changes our results. In practice we chose $\omega_{\infty}=1500 \pi T$, as further increasing the maximum frequency did not lead to major changes in self-energy or critical scales. 

This way we calculate the vertex and self-energy only for a given set of Matsubara frequencies and momentum vectors on the Fermi surface. The flowing self-energy then appears on the internal lines of the loop diagrams, as visible in Eqs. (\ref{loops}) and (\ref{singlescale}). Importantly, we neglect the real part of the self-energy. This is because a feedback of the real part of the self-energy would lead to a Fermi surface shift which is hard to handle within the $N$-patch scheme, as the Fermi surface could be shifted into a momentum region which has already been integrated out in previous fRG steps, causing severe singularities in the flow.

\section{\label{results}Results}
\subsection{Case without instability}
\begin{figure}
\begin{center}
\includegraphics[width=.48\linewidth]{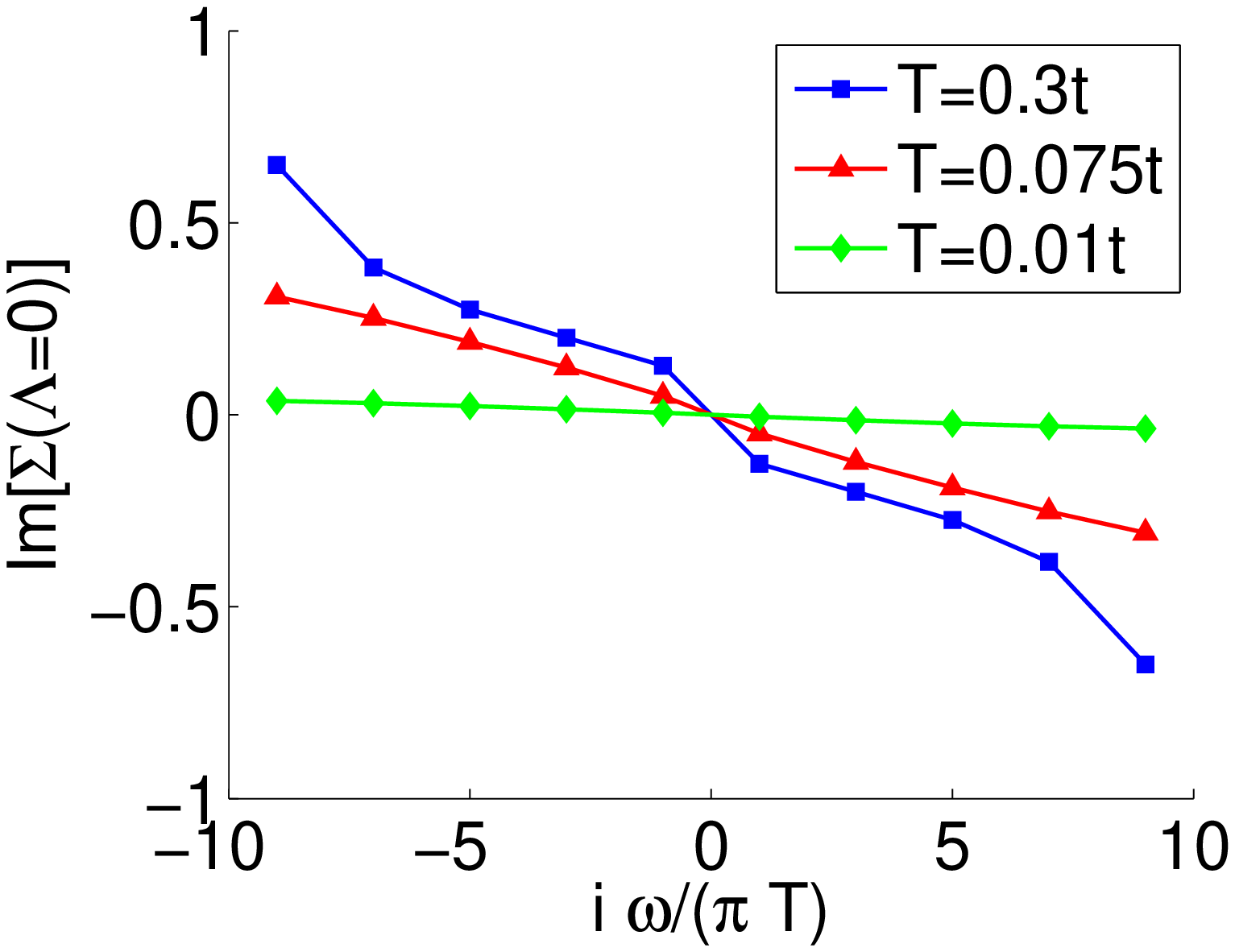}
\includegraphics[width=.48\linewidth]{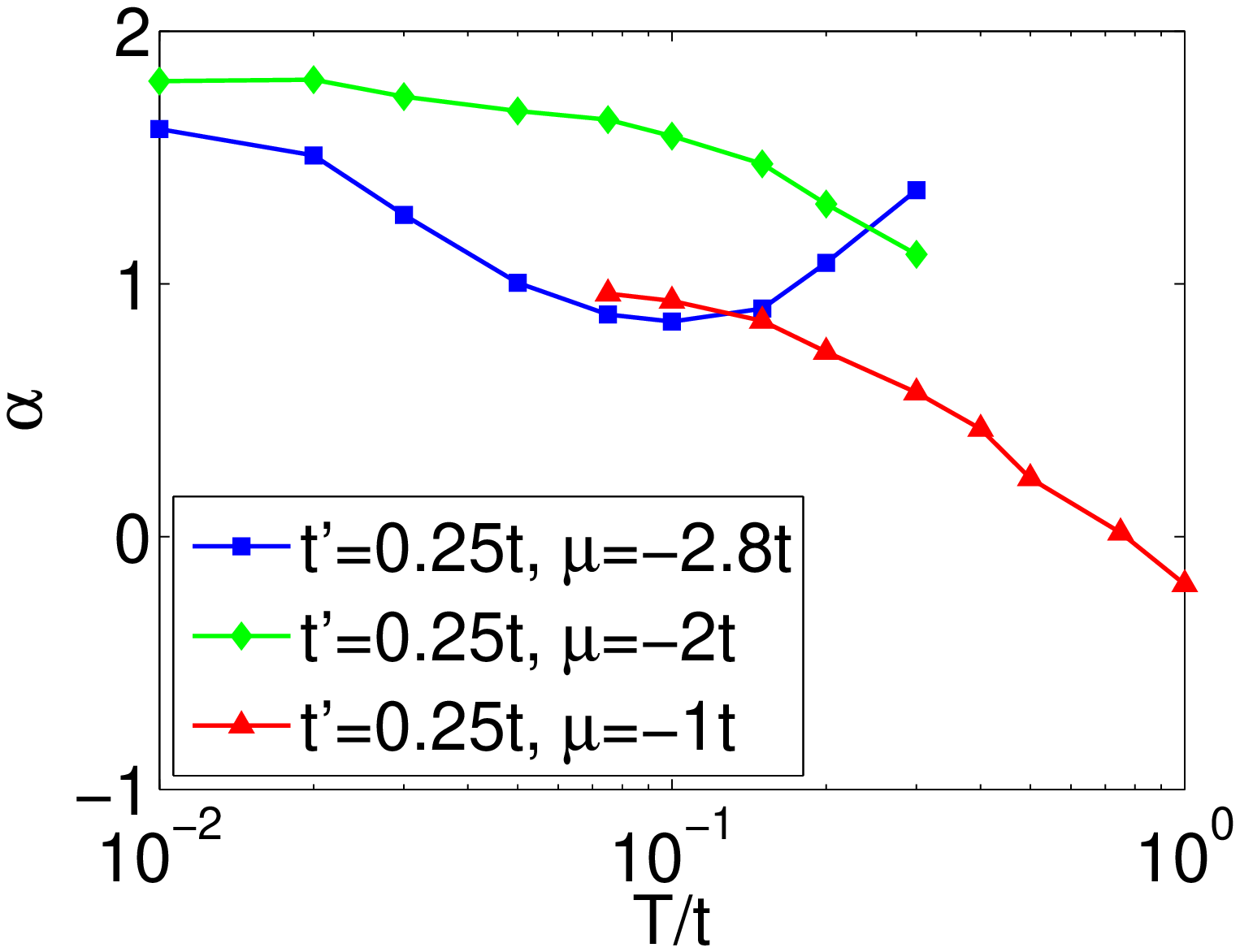}
\end{center}
\caption{Left: Imaginary part of the self-energy in units of $t$ for different $T>T_c$. $t'=0.25t$, $\mu=-2t$. Shown are the values at momentum $\mathbf{k}=1$ (see Fig. \ref{patchingplot}). The largest couplings at $\Lambda=0$ are of the order of $4t$. 
Right: Estimation for the exponent in Eq. (\ref{tau-alpha}) for $t'=0.25$ at van Hove filling (red triangles) and away from van Hove filling (blue squares, green diamonds), obtained via Eqs. (\ref{deltasigma}) and (\ref{alpha-ti}) for $\mathbf{k}=1$ (see Fig. \ref{patchingplot}). All data for $U=3t$ with self-energy feedback in the discretization D1.}
\label{plot-novh} 
\end{figure}

First, as a sanity check in order to get familiar with the data we obtain from the fRG, let us look at a case where the flow indicates normal metallic behavior. For that we choose $t'=0.25t$ and $\mu= -2t$ such that the Fermi surface is away from the van Hove situation and from the antiferromagnetic (AF) spin-density wave (SDW) instability. Then, if we stay away from lowest temperatures, the fRG flow does not lead to strong coupling, i.e. we can integrate out all modes without encountering a strong growth of any components of the coupling function. In Fig. \ref{plot-novh} we show the imaginary parts of the self-energy as function of the Matsubara index for different temperatures. Besides a decrease of the self-energy toward lower $T$, we observe a monotonic frequency-dependence that features some kind of downward step around zero frequency. The slope of the overall curve can easily be interpreted as quasiparticle weight or $Z$-factor, while the step can be understood as a finite lifetime $\tau$.

Note that for higher frequencies of the scale of the bandwidth, the imaginary part of the self-energy should approach zero again, but our frequency window is not large enough to capture this behavior. In our treatment, the self-energy at frequencies outside the discretization window (i.e. for Matsubara indices higher than $\pm 10$) is approximated by the value at the closest discretization frequency. This may underestimate the true self-energy in the frequency range where the linear slope around zero frequency still continues, but may overestimate it at higher frequencies where the imaginary part of the self-energy is a decreasing function of the absolute value of the frequency. To some extent these two effects will cancel each other, and the large-frequency self energy is anyway less relevant for the Fermi surface instabilities considered here. Therefore we think that our simple scheme of treating the self-energy behavior at higher frequencies is qualitatively adequate.   

We can analyze the behavior of the quasiparticle lifetime obtained from the step around zero frequency a bit more closely.  In a Fermi liquid, the life-time is related to the temperature as $\tau^{-1} \sim T^2$, with a logarithmic correction in two dimensions that we might not be able to resolve. Let us test if the law
\begin{equation}
\label{tau-alpha}
\tau^{-1} \propto T^\alpha
\end{equation}
is verified by our data, where we should find $\alpha=2$ for a Fermi liquid with round Fermi surface. 
At van Hove filling due to the divergence of the density of states the exponent is expected to be altered to $\alpha=1$ \cite{lee}. Note however that all these literature values are obtained by finite-order perturbation theory, while the fRG sums up infinite orders in the bare coupling.
The fRG estimate for the life-time is obtained by linearly extrapolating the self-energy for the two lowest frequencies to $\omega=0$
\begin{equation}
\label{deltasigma}
\tau^{-1}=\Delta \Sigma^\Lambda=\frac{1}{2} \text{Im}[\Sigma^\Lambda(\mathbf{k}_F, 3 \pi T)] - \frac{3}{2}\text{Im}[ \Sigma^\Lambda(\mathbf{k}_F, \pi T)]
\end{equation}
The expression
\begin{equation}
\label{alpha-ti}
\alpha_{T_i}=\frac{\Delta(\log(\tau^{-1}))}{\Delta(\log(T))}= \frac{\log(\tau^{-1}_{T_{i+1}})-\log(\tau^{-1}_{T_{i}}) }{\log(T_{i+1})-\log(T_i)}
\end{equation}
gives an estimate for the exponent, extracted from data at two temperatures $T_i$ and $T_{i+1}$. The right plot of Fig. \ref{plot-novh} shows the temperature-dependence of $\alpha$. Although we do not observe a constant $\alpha$ presumably due to the error of the extrapolation, at least $\alpha$ for the rounded Fermi surface is near the expected value of two for low enough temperatures, and similarly for van Hove filling $\alpha$ is close to one for low temperatures. 

\subsection{Critical scales and leading instabilities}
Now let us come to cases with flows to strong coupling.  The first question we want to ask is whether the inclusion of frequency-dependence and feedback from the imaginary part of the self-energy leads to a different leading instability or a change in the critical scale. Here we study this question along the parameter line of van Hove fillings where the Fermi surface contains the points $(\pi , 0)$ and $(0, \pi )$. These are saddle points of the dispersion and cause a divergence of the density of states. From previous works it is known that this van Hove situation leads to competing ordering tendencies. Hence this situation is a good arena to study the impact of the inclusion of frequency-dependence and self-energy feedback.

First let us look at the van Hove situation at half filling, where in addition the Fermi surface is perfectly nested. Previous fRG works have clearly established an AFM instability, signaling an AF SDW ground state for this case. This in in accordance with the expectations from simple RPA arguments, but also with quantum Monte Carlo \cite{varney}.

\begin{figure}
\begin{center}
\includegraphics[scale=0.4]{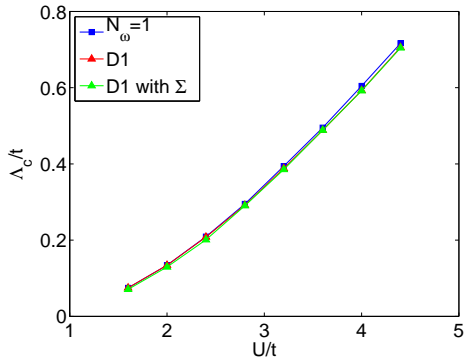}
\end{center}
\caption{Critical scale, at which the couplings exceed $20t$, vs. $U$ for the fully nested case, with $T=0.01t$, $t'=0$, and $\mu=0$ for the flows with frequency independent vertices  without self- energy feedback (blue squares),  with frequency-dependent vertices and without self-energy feedback (red triangles), with frequency-dependent vertices and with self-energy feedback (green diamonds).}
\label{uvglplot} 
\end{figure}

For $t'=0$, $\mu=0$, the character of the instability in the fRG does not change when we include a frequency-dependence of the effective interactions, or when we include the self-energy feedback. This means that the couplings that drive the antiferromagnetic susceptibility with wave vector transfer $(\pi,\pi)$ diverge most strongly, as described in earlier works\cite{metznerRMP}. 
In Fig. \ref{uvglplot} we show the critical scales obtained in the fRG in the various approximations (without frequency-dependence of the effective interactions and without self-energy feedback, with frequency-dependence of the effective interactions but without self-energy feedback, with frequency-dependence of the effective interactions and with self-energy feedback) as function of the interaction strength $U$. These scales do not differ much quantitatively. One way to understand the agreement is that the instability in this case is very clue to a standard RPA instability, where only the most singular frequency transfer matters, and different frequency transfers do not couple, hence the frequency resolution does not come into play. 
Furthermore, the flow goes off to strong coupling at scales before the self-energy becomes noticeable.  Neverthelss, with respect to plain RPA, the critical scales are reduced by the channel coupling in the fRG.
Hence, in this case, the simplest fRG approximation without frequency-dependence and without self-energy feedback is already quite good and cannot be improved much within the range of numerical possibilities.   The remaining approximations are besides the discretizations of wave vector and frequency-dependences the neglect of the real part of the self-energy and the truncation of the flow hierarchy after the four-point function. We expect that the any remaining differences are due to these two approximations. In particular for larger $U \sim 4t$, there should be precursors of the spectral weight transfer that ultimately leads to the opening of a Mott gap at $U \sim 8t$. As is believed that the truncated fRG does not allow the description on the Mott transition, it is plausible that the spectral weight transfer is still not described correctly in our improved approximation. Furthermore, the physics of collective fluctuations might not be captured to some extent due to the truncation after the four-point vertex.
Hence, if we compare roughly to gap scales found in non-perturbative cluster calculations for large clusters\cite{moukouri}, the $\Lambda_c$-curve seems to rise too steeply as function of $U$. However, regarding the critical scale as estimate of the gap scale, up to a factor of order one, the fRG is in the right range. 

\begin{figure}
\begin{center}
\includegraphics[width=.48\linewidth]{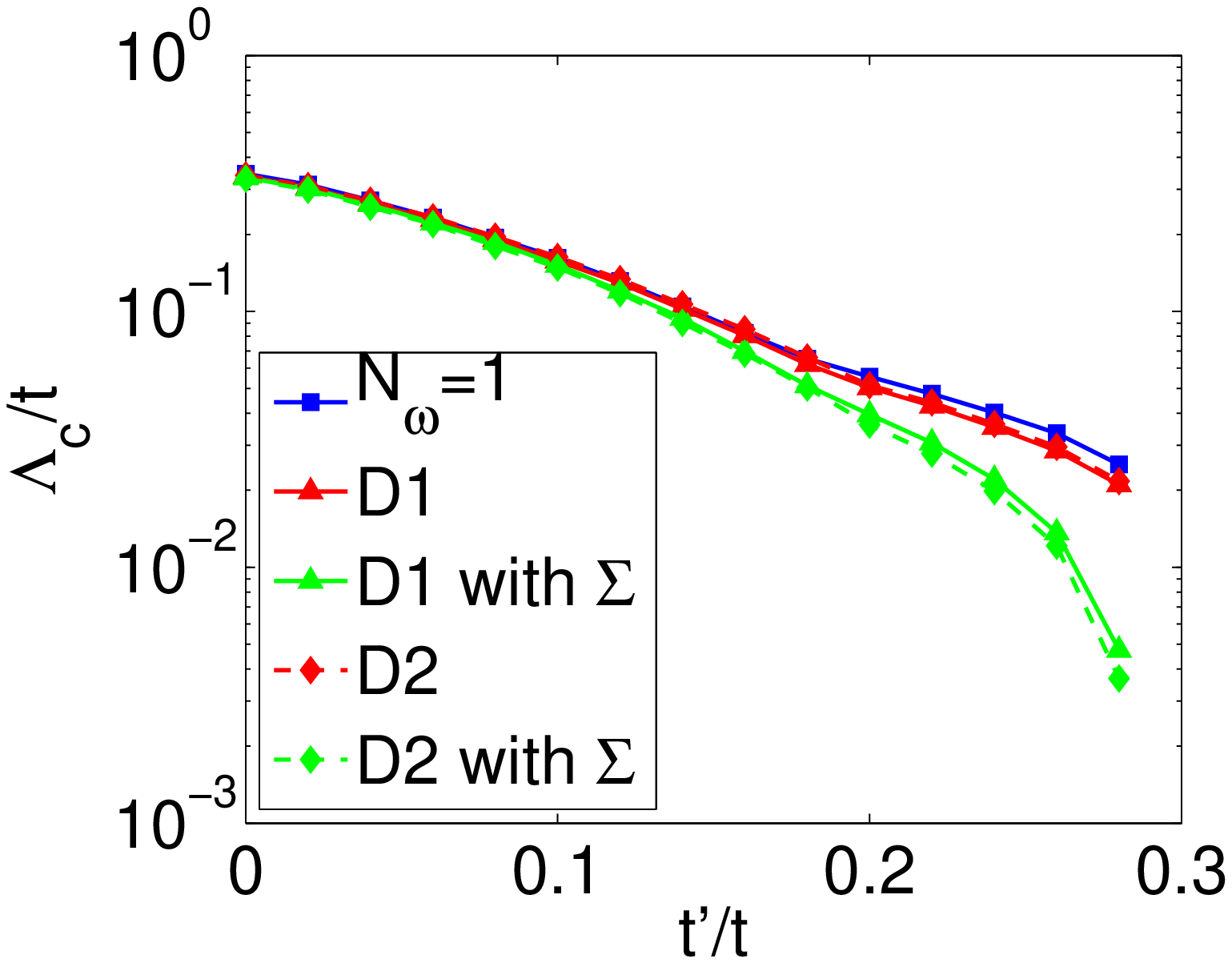}
\includegraphics[width=.48\linewidth]{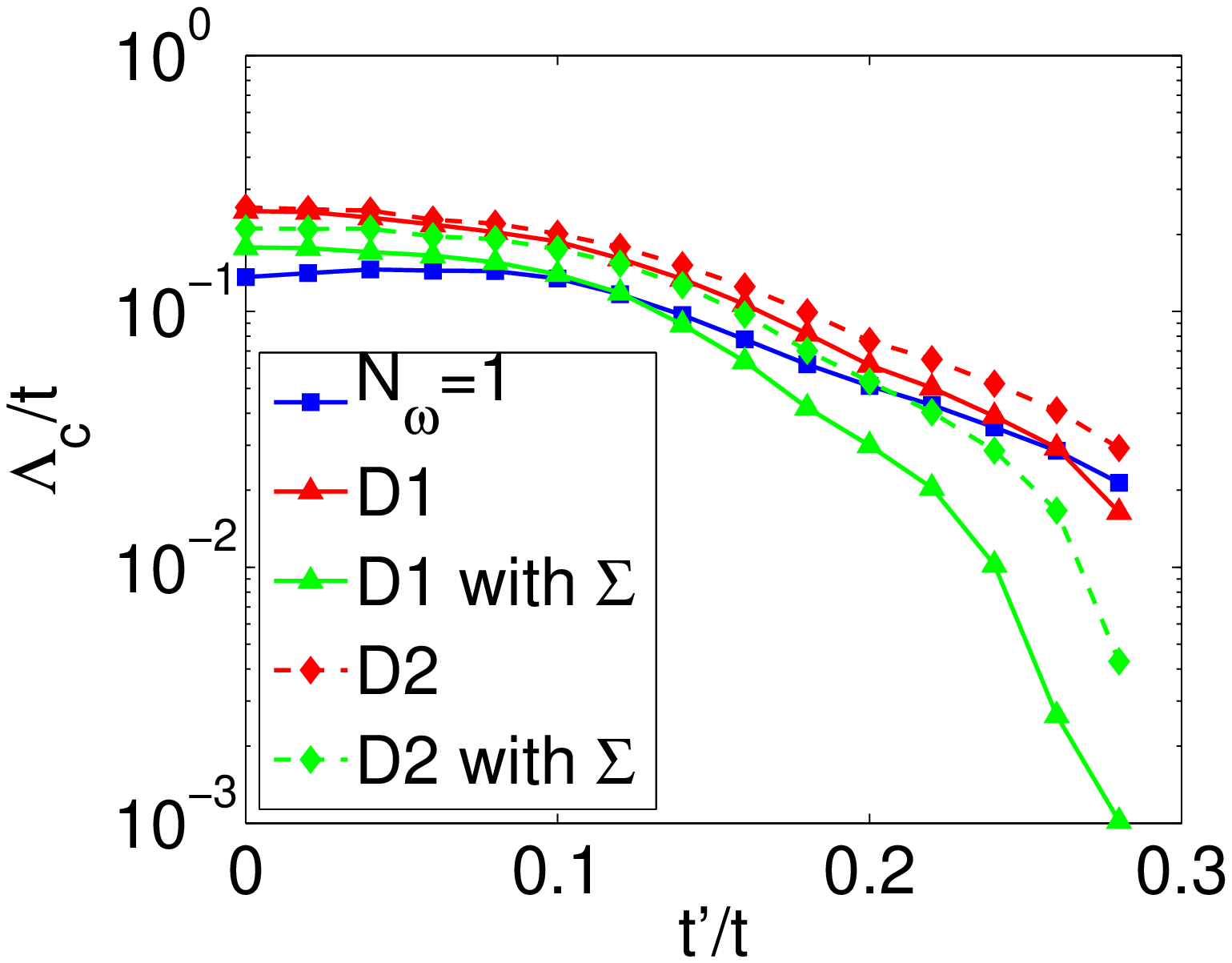}
\end{center}
\caption{Critical scale $\Lambda_c$, at which the couplings exceed $20t$, vs. next-nearest-neighbor hopping $t'/t$ at van Hove filling for $U=3t$ with $N_\omega=1$ and without self- energy feedback (blue squares),  with frequency-dependent vertices and without self-energy feedback (red), with frequency-dependent vertices and with self-energy feedback (green). Two different frequency discretization D1 (triangles, solid lines) and D2 (diamonds, dashed line) are used. Left: $T=0.01t$. Right:  $T=\Lambda_c^{0.01}$ is the the critical scale for $T=0.01t$ without self-energy feedback with discretization D1 (red solid line in the left plot).}
\label{tsvglplot} 
\end{figure}

Let us now look at the situation where the particle-hole nesting is destroyed by larger $t'$. In Fig \ref{tsvglplot}, we  plot the critical scale vs. the next-nearest-neighbor hopping $t^\prime$ at van Hove filling (i.e. chemical potential $\mu=-4t^\prime$), at which the van Hove energy with a logarithmic diverging density of states is at the Fermi level. The fRG flows in Fig \ref{tsvglplot} a) have been computed at $T=0.01t$. Lowering the temperature further increases the numerical effort, as the maximum cutoff frequency $\omega_\infty$ then has to be taken larger too. 

As can be seen in Fig. \ref{tsvglplot} a) for $T=0.01t$ the inclusion of frequency-dependence and the different discretizations affects the critical scale only slightly. The self-energy feedback however reduces the critical scale significantly especially for larger $t^\prime$.  We also compared how different patchings on the Matsubara axis affects the critical scales. In Fig. \ref{tsvglplot} we show data for two different discretizations of the Matsubara axis. At low temperatures, the two different discretization lead to very similar results, while at higher temperatures the two  discretizations show some quantitative differences, depending on $t'$.

One might suspect that the critical scale for higher $t'$ is more strongly suppressed because the critical scale comes closer to the chosen temperature, so that the difference in the critical scale is merely a finite temperature effect. To check this we plot the critical scale for different $t'$ with $T=\Lambda_c^{0.01}$, where $\Lambda_c^{0.01}$ is the critical scale at $T=0.01t$ without self-energy feedback. Now the ratio of temperature and critical scale is comparable for all $t'$. We see that inclusion of self-energy-feedback suppresses the critical scale now already for small $t'$, but definitely more strongly at larger values of $t'$. 

Interestingly we find that increasing the temperature reduces the critical scale more strongly, when frequency-dependence is neglected at small $t'$, so that we discover a parameter space, in which now the critical scale increases when one includes frequency-dependence in contrast to Fig. \ref{tsvglplot} a). At larger temperatures the difference between the two discretization schemes also becomes more pronounced. The flows with the discretization D2 now exhibit a larger critical scale than the flows with D1. Hence the precision of the description is better at low $T$.

\begin{figure}
\begin{center}
\includegraphics[width=.48\linewidth]{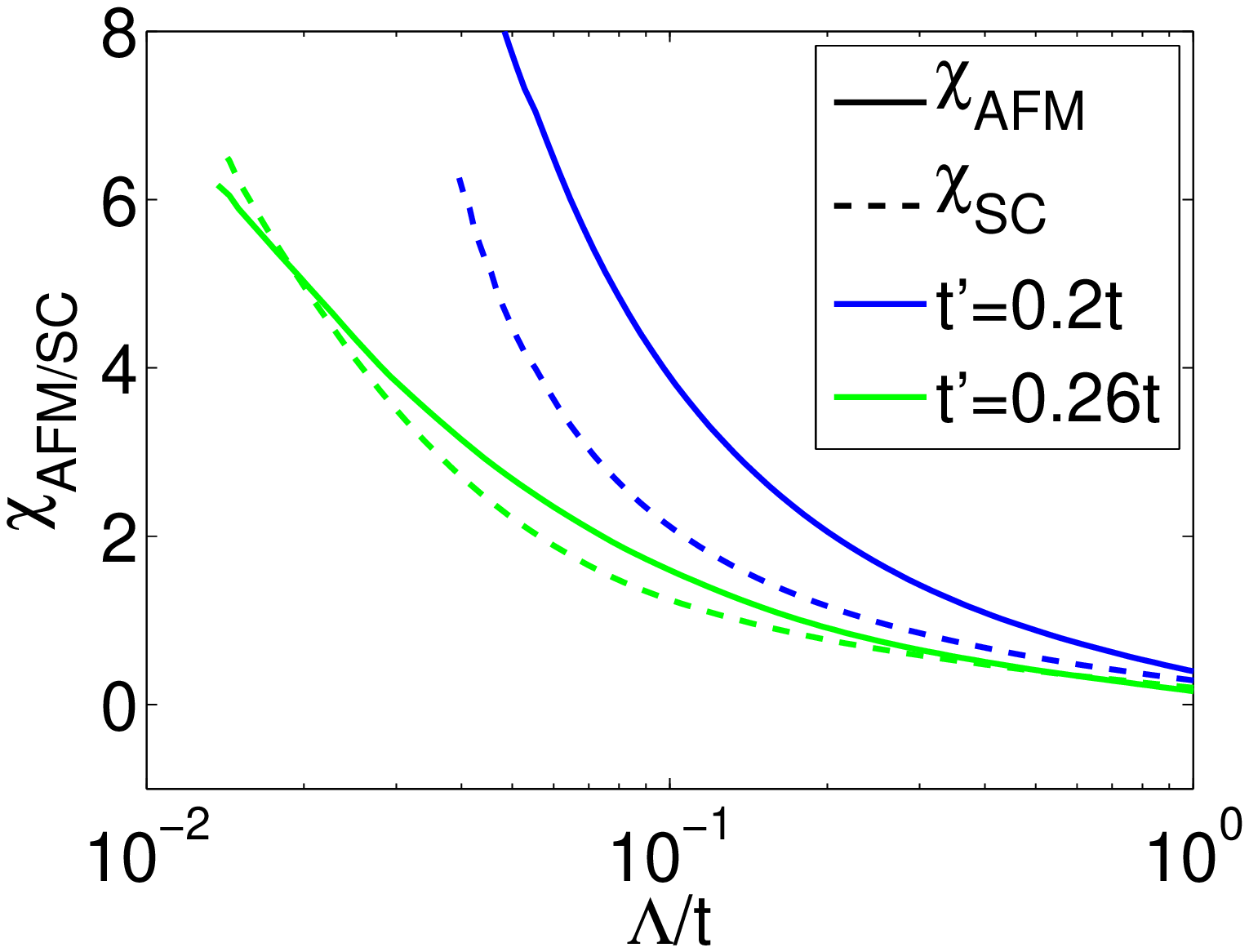}
\includegraphics[width=.48\linewidth]{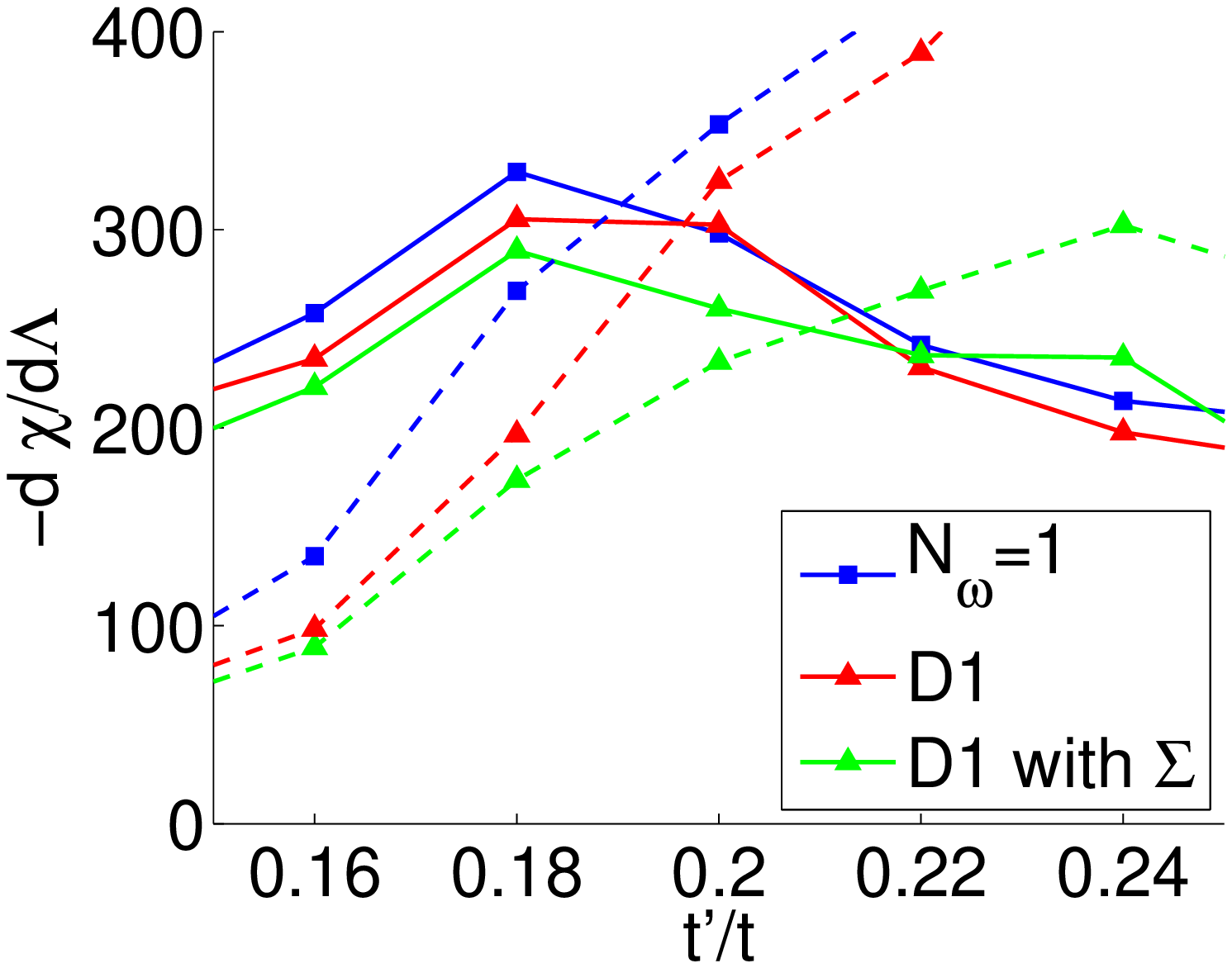}
\end{center}
\caption{Left: Flow of the effective strength of the antiferromagnetic (solid lines) and superconducting (dashed lines) channel given in Eq. (\ref{susafm}, \ref{sussc}) in units of $t$  for $t'=0.2t$ (blue) and $t'=0.26t$ (green) in the discretization scheme D1 and with self-energy feedback. Right: Slope of the effective strength of the antiferromagnetic (solid lines) and superconducting (dashed lines) channel at the scale $\Lambda_c$, at which the couplings exceed $20t$. Shown are the data with frequency independent vertices and without self- energy feedback (blue squares),  without self-energy feedback (red triangles), and with self-energy feedback (green triangles), both in the discretization scheme D1. The intersection point of the two curves is used as an estimate for the boundary between the AFM and SC dominated regime. All data with $U=3t$, at van Hove filling and $T=0.01t$.}
\label{fig-regimes} 
\end{figure}

We want to have a closer look at the origin of the behavior observed in Fif. \ref{tsvglplot}. A natural question to ask is whether the AFM SDW and the superconducting (SC) tendencies respond differently to the inclusion of frequency-dependence and self-energy, and whether the tentative phase diagrams drawn by determining the leading sintability are changed by these improvements of the scheme.
We therefore consider the effective coupling strengths of the AFM and SC channel, which we define as  

\begin{equation}
\label{susafm}
\chi_{\text{AFM}}=\frac{1}{N_k}\sum_{\mathbf{p}}
V^\Lambda \left( (\mathbf{p},\omega_0),(\mathbf{k}=1 ,\omega_0), (\mathbf{k}=24,\omega_0)  \right), \\
\end{equation}

\begin{equation}
\label{sussc}
\chi_{\text{SC}}=\frac{1}{N_k^2}\sum_{\mathbf{k},\mathbf{p}}
V^\Lambda \left( (\mathbf{k},\omega_0),(-\mathbf{k},-\omega_0),(\mathbf{p},\omega_0)  \right) f(\mathbf{k}) f(\mathbf{p}) \, 
\end{equation}      
Here, $\mathbf{k}=1$ and $\mathbf{k}=24$ denote the patches with momentum vectors closest to the van Hove singularities roughly connected by momentum transfer $(\pi,\pi)$ (see Fig. \ref{patchingplot}) and $\omega_0=\pi T$. $f(\mathbf{k})$ and $f(\mathbf{p})$ are $d$-wave form factors and are given by $f(\mathbf{k})=\cos(k_{x})-\cos(k_{y})$. We consider the vertices of the smallest Matsubara frequencies, as these grow most strongly during the fRG flow. If these averages diverge, the corresponding susceptibilities diverge as well, so that these quantities can be used as a measure of the coupling strengths of the respective channel. Typical flows of these quantities are shown in the left plot of Fig. \ref{fig-regimes}.  As a criterium for the leading instability, we choose the derivatives of the coupling strength of the effective channels with respect to $\Lambda$ at $\Lambda_c$. $\Lambda_c$ is again defined as the scale at which the couplings exceed a value of $20t$. The results are shown in the right plot of Fig. \ref{fig-regimes}. The intersection of the respective curves for the AFM and SC coupling strength is taken as an estimate for the boundary between the regimes. Of course the precise values from this procedure depend somewhat on the chosen definition of $\Lambda_c$. On the other hand, this procedure suffices for the more qualitative discussion, which we are interested in, namely the impact of frequency-dependence and self-energy effects on the competition between the different orders. The comparison for the different cases, i.e. with and without frequency-dependence and self-energy feedback, is summarized in table \ref{tab-regimes}. 

\begin{table}
\begin{tabular}{c|c|c|c|c|c}
 	&  \multicolumn{3}{c}{without $\Sigma$}  &	\multicolumn{2}{|c}{with $\Sigma$}	\\ %\cline{2-6}
	& $N_\omega=1$ 			& D1		&	D2		  &   D1	&	D2			\\ \hline
  $T=0.01t$ &		$0.19$	&  $0.2$	&	$0.21$		  &	$0.21$	&	$0.22$			\\
$T=\Lambda_c^{0.01}$&	$0.22$	&  $0.27$	&	$\gtrsim 0.28$		  &	$>0.28$	&	$>0.28$			\\
\end{tabular}
\caption{Approximate values for $t'/t$, for the transition from the AFM to the SC regime at van Hove filling and $U=3t$. The values are obtained by the comparison of the slope of the effective strength of the respective channels at the scale $\Lambda_c$, at which the couplings exceed $20t$ (see also Fig. \ref{fig-regimes}).}
\label{tab-regimes}
\end{table}

At $T=0.01t$ and without self-energy the boundary between the regimes lies at $t' \approx 0.19t$ and without frequency-dependent vertices. If we include the feedback of self-energy this boundary is shifted towards higher $t'$. We conclude that life-time effects harm the superconducting instability more than the AFM instability. 

For the second run at higher temperatures $T= \Lambda_c^{0.01}$ the phase boundary is generally shifted towards higher $t'$, too. This basically reflects the fact that the relevant energy scale of the superconducting instability is lower than the one of the AFM instability. Hence the pairing channel is more strongly affected by the finite temperature. The difference between the different levels of approximations is much larger than at $T=0.01t$ in correspondence with the larger discrepancy in the critical scales in Fig. \ref{tsvglplot}. 
Furthermore, we note that the different discretization schemes yield qualitatively similar results, which confirms that the picture drawn here for the leading instabilities is fortunately rather independent on details of the numerical implementation.
 
The overall conclusion we can draw from this analysis is that the frequency-dependence of the couplings and the self-energy effects are certainly important for quantitative questions. However, they in general do not qualitatively change the physical picture obtained from the simpler flow schemes with frequency-independent couplings and with neglected self-energy.

\begin{figure}
\begin{center}
\includegraphics[width=.48\linewidth]{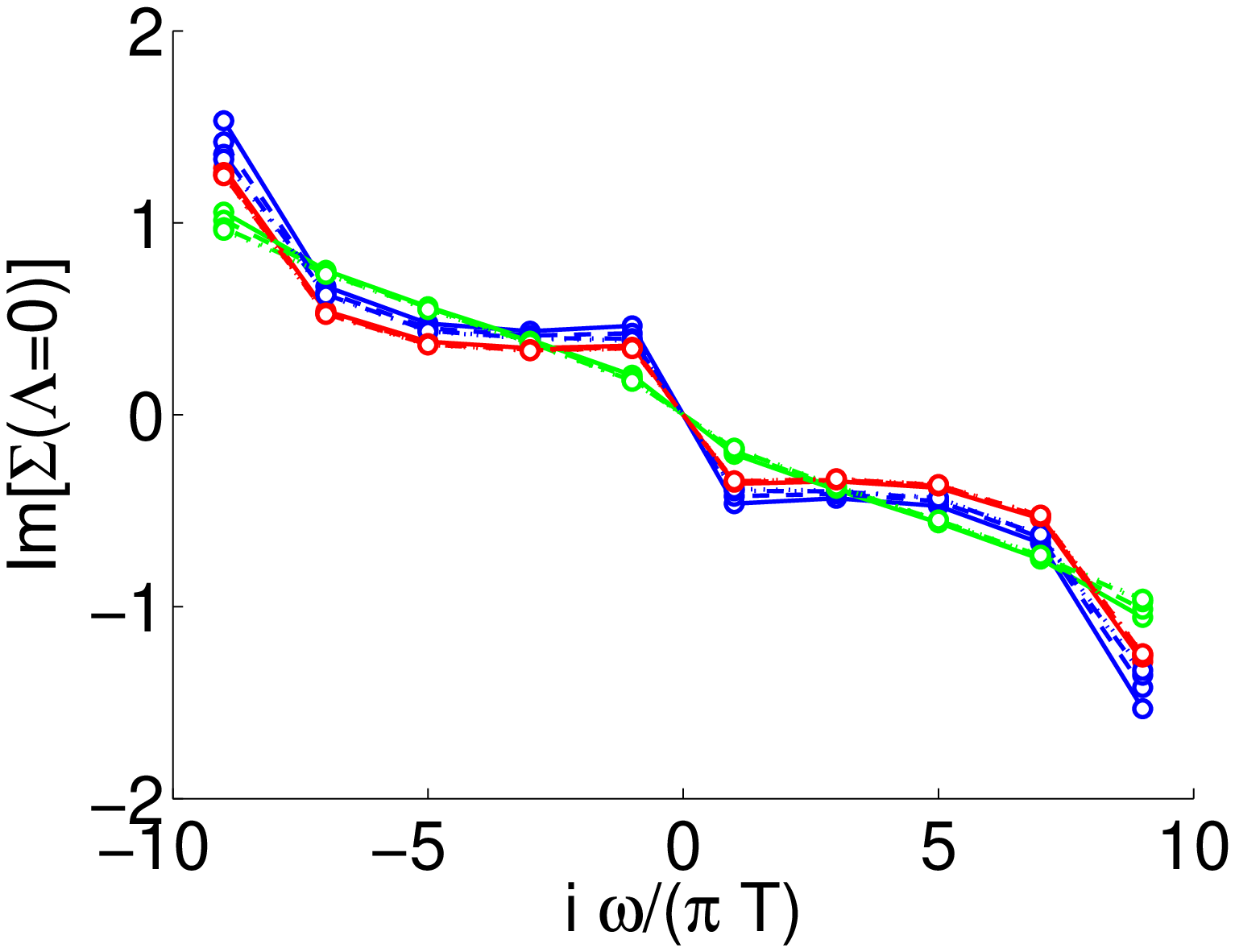}
\includegraphics[width=.48\linewidth]{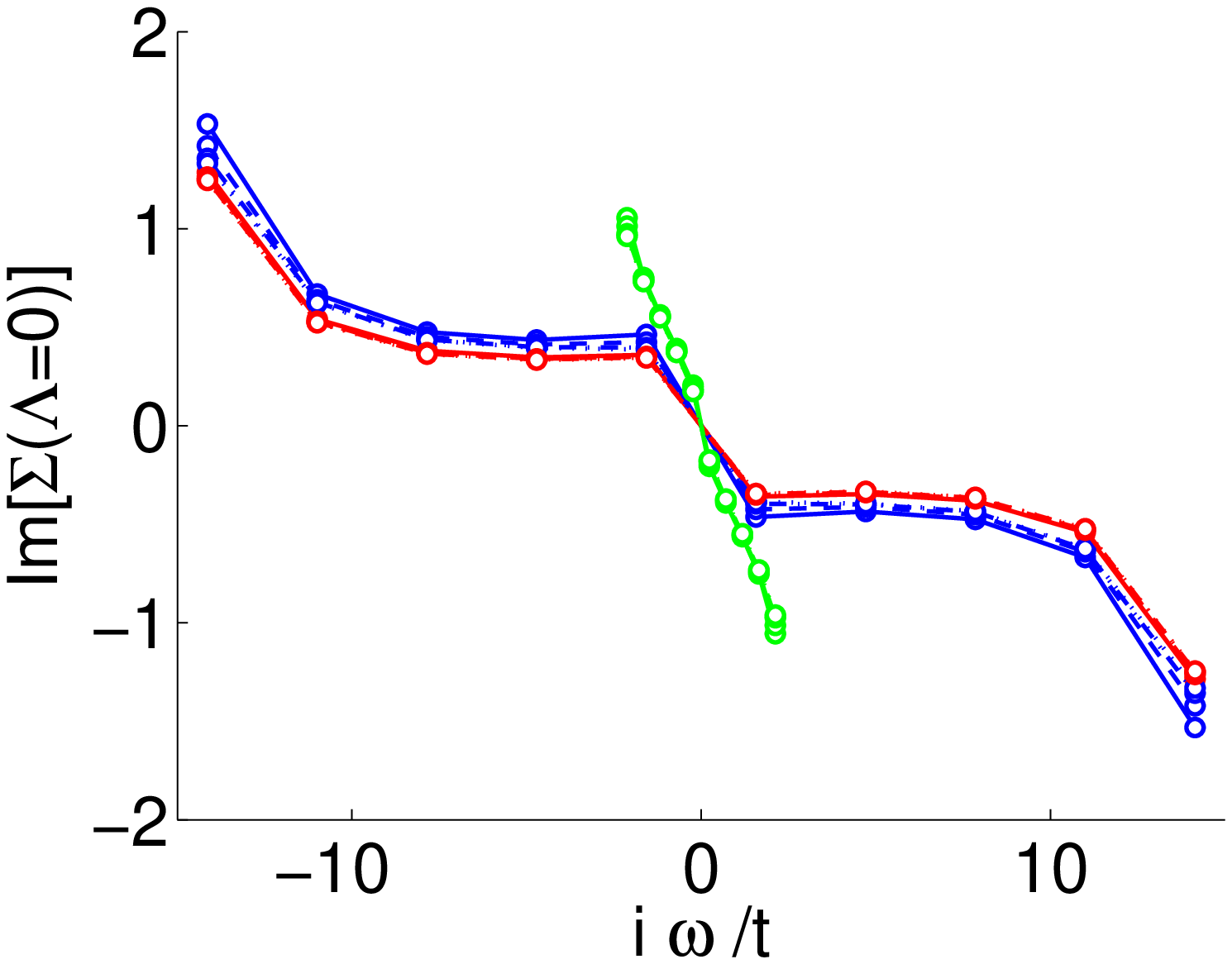}
\end{center}
\caption{Imaginary part of the self-energy in units of $t$ for $T>T_c$ at $\Lambda=0$ for three parameter choices at van Hove filling, $U=3t$. $T=0.5t$, $t'=0$ (blue);  $T=0.5t$, $t'=0.25t$ (red); $T=0.1t$, $t'=0.25t$ (green). Shown are different patches, i.e. momenta at the Fermi surface $\mathbf{k}=1$ (solid line), $\mathbf{k}=2$ (dashed line), $\mathbf{k}=3$ (dashed dotted line), $\mathbf{k}=4$ (dotted line), all nearly on top of each other: only weak angular dependence. The left and right plot shows the same data with different scaling of the frequency axis. All data with discretization scheme D1 and feedback of self-energy. The largest couplings at $\Lambda=0$ are of the order of $10t$ for $T=0.5t$, $t'=0.25t$ (red). The parameter choices shown in blue and green are close to the critical temperature, here the largest couplings are close to $20t$. }
\label{phiplot} 
\end{figure}

\subsection{Flow of the self-energy}
Next we consider the flow of the self-energy for finite temperatures above $T_c$. The main goal is now to analyze the frequency- and wavevector-dependence.

We run the fRG flow for different $T>T_c$ such that all modes can be integrated out. Practically, the flow is stopped at a scale $\Lambda=2 \times 10^{-4}t$, which is much lower than the temperature. Below these scales, the vertices and self-energies do not get renormalized substantially by lowering $\Lambda$ even further. We stay in the weakly to moderately coupled regime at all scales. 

In Fig. \ref{phiplot} the imaginary part of self-energy is plotted over the Matsubara frequency for different momentum patches, for the fully nested case with $t'=0$ at half filling and for van Hove filling with $t'=0.25t$. There is only a weak angular dependence at these elevated temperatures above the instability, independent of the Fermi surface shape. We checked that the weak angular dependence is reproduced in the other frequency discretization schemes as well.
On the left hand side of the plot, we show the data as function of the Matsubara frequency, i.e. with different frequency window $\propto T$ for the two temperatures. On the right hand side we show the same data vs. $i\omega/(\pi T)$ as function of the Matsubara frequency  index.

\begin{figure}
\begin{center}
\includegraphics[scale=0.4]{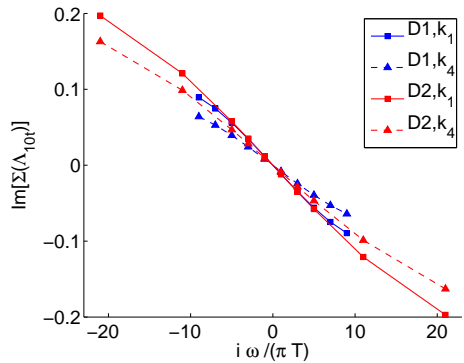}
\end{center}
\caption{Imaginary part of the self-energy in units of $t$ for different discretization schemes D1, D2 at the scale $\Lambda_{10t}$ at which the couplings exceed a value of $10t$. $\Lambda_{10t}=0.084t$ ($\Lambda_{10t}=0.093t$) for D1 (D2). Shown are different momenta at the Fermi surface $\mathbf{k}=1$ and $\mathbf{k}=4$ (see Fig. \ref{patchingplot}). All data for $T=0.01t$, $t'=0.2t$, $U=3$, at van Hove filling and with feedback of self-energy.}
\label{Tplot} 
\end{figure}

At higher temperatures $T=0.5t$, we observe again a step-like discontinuity in the self-energy at zero frequency on the Matsubara axis. This can be interpreted as an inverse lifetime of the quasiparticle peak. If we go to lower temperatures but remain above the instability (now only for the curved Fermi surface, as otherwise the critical scale is too high), this step gets much smaller (also because the Matsubara frequencies move close together) and we are left with a rather linear frequency-dependence, which can again be captured by a quasiparticle weight $Z$. This again turns out to be rather independent of the location on the Fermi surface. So all this data at fixed temperature looks quite consistent with a rather normal and isotropic metallic system, at least at these temperatures above the instability.

There have been a number of more detailed studies (e.g. Refs. \onlinecite{feldman,gonzalez}) of the self-energy in the case of van Hove filling, pointing out special, non-Fermi liquid-like properties due to the diverging density of states. At a given $T$, it is difficult to read out any particular self-energy property from our data at $t'=0.25t$. However, if we measure the temperature-dependence of the step-like feature that we interpreted as inverse lifetime and fit it to the law $T^\alpha$, we obtain a different exponent with $\alpha $ approaching 1 toward lower $T$. This marginal Fermi-liquid behavior\cite{varmaMFL} is shown in Fig. \ref{plot-novh} and consistent with second-order predictions for the van-Hove situation\cite{lee,hlubina}.

In Fig. \ref{Tplot} we show data lower temperatures for the frequency-dependence of the self-energy, now comparing different frequency discretizations. We compare data at a scale where the couplings exceed a value $10t$. Note that this is at slightly different scales in both cases. We see that for low frequencies the discretizations compare quite well. At larger frequencies the quantitative agreement becomes worse. However, the qualitative features, as for instance a larger slope of the self-energy near the saddle points compared to the Brillouin zone diagonal, is found consistently in both schemes. Quite generally, our data shows that the self-energies in the two schemes are more consistent at lower temperatures. 

\begin{figure}
\begin{center}
\includegraphics[scale=0.4]{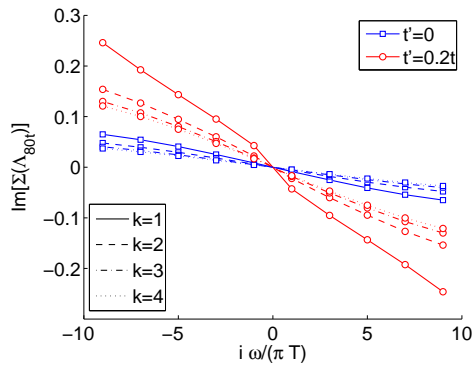}
\end{center}
\caption{Imaginary part of the self-energy in units of $t$ for $t'=0$ (blue) and $t'=0.2$ (red) at van Hove filling for $U=3t$ at the scale $\Lambda_{80t}$ where the couplings exceed a value of $80t$. $\Lambda_{80t}=0.26t$ ($\Lambda_{80t}=0.019t$) for $t'=0$ ($t'=0.2t$). Shown are different momenta at the Fermi surface $\mathbf{k}=1$ (solid line), $\mathbf{k}=2$ (dashed line), $\mathbf{k}=3$ (dashed dotted line), $\mathbf{k}=4$ (dotted line) (see Fig. \ref{patchingplot}). All data with discretization scheme D1 and with feedback of self-energy.} 
\label{hvmplot} 
\end{figure}

For $T<T_c$ the flows go to strong coupling. If we stop the flow when the vertices exceed a value of the bandwidth, the momentum-dependence of the self-energy are rather weak. However, if we continue to flow towards the instability, stronger anisotropies are found. 
Strictly speaking, for these large values of the vertices the truncation of the fRG flow equation with neglect of the higher-order vertices is no longer justified. We expect however that we can still get some qualitative insight in the breakdown of the Fermi liquid behavior for low temperatures, similar to the argumentation in previous works\cite{kataninkampf,rohe}.
The corresponding data is shown in Fig. \ref{hvmplot}, both with and without self-energy feedback.

\begin{figure}
\begin{center}
\includegraphics[width=.48\linewidth]{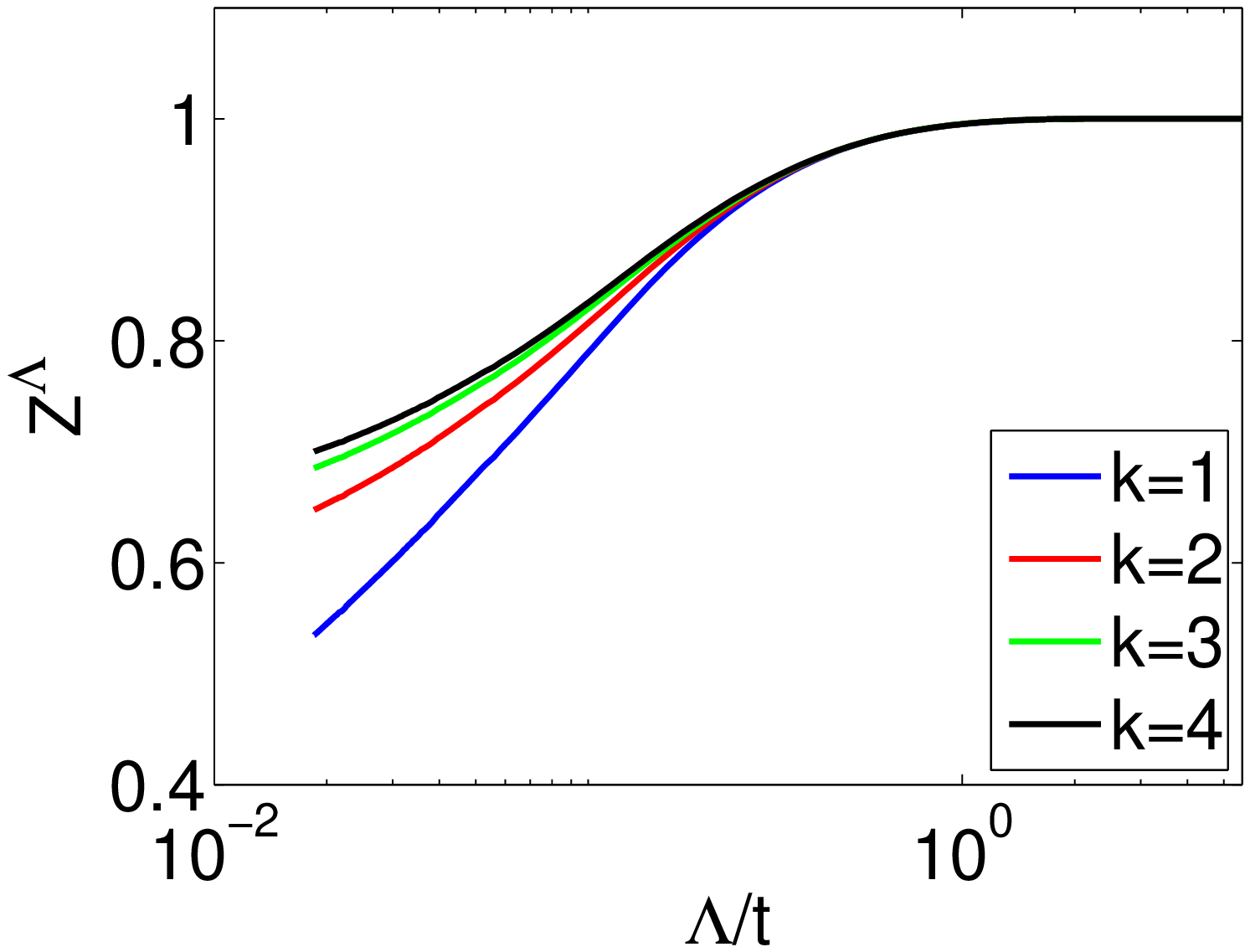}
\includegraphics[width=.48\linewidth]{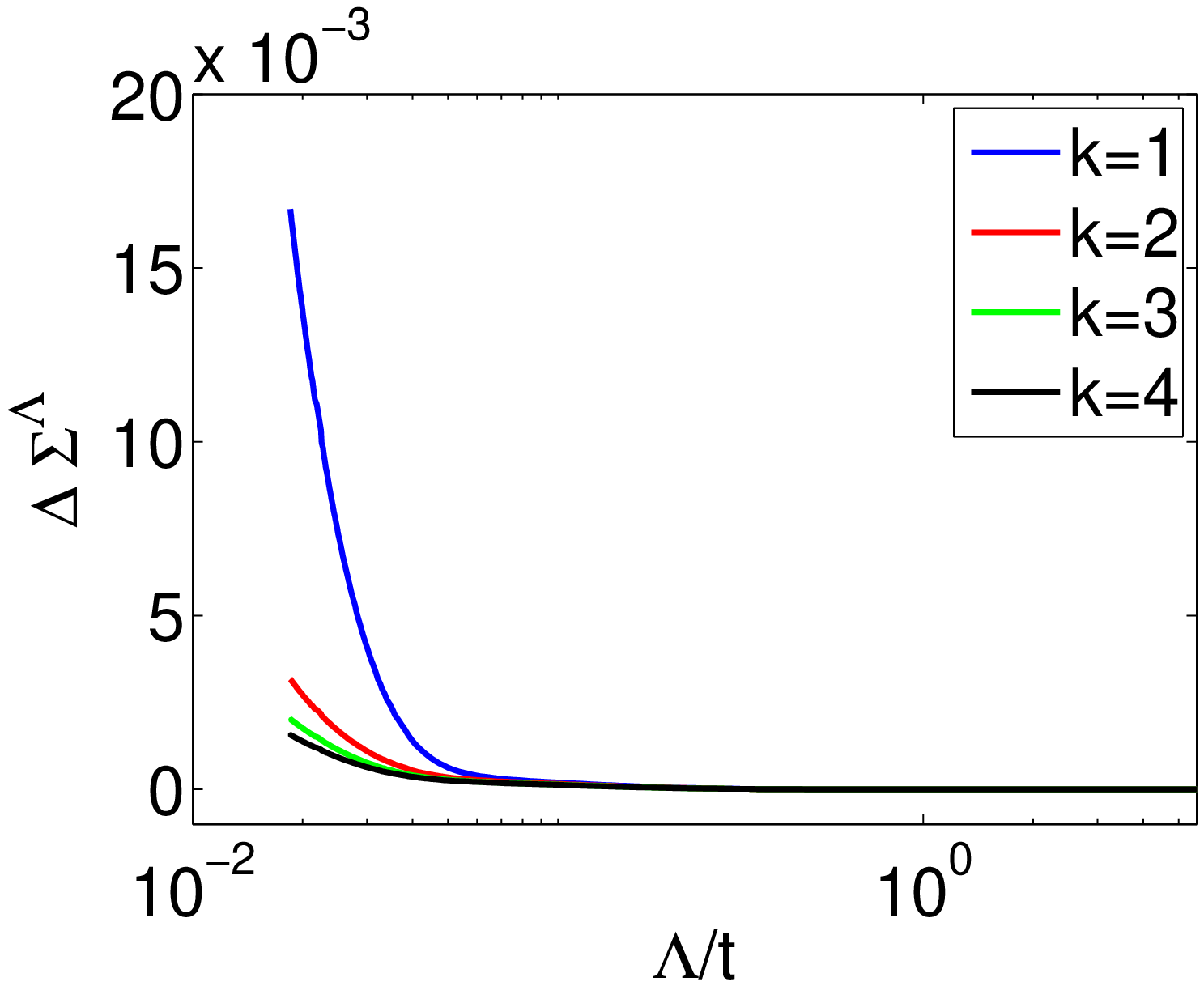}
\end{center}
\caption{Flow of the $Z$-factor as given by Eqs. (\ref{zfactor}, \ref{dsigma-domega}) and the step $\Delta \Sigma^\Lambda$ of the self-energy in units of $t$ at $\omega=0$ defined as in Eq. (\ref{deltasigma}) for different momenta at the Fermi surface (see Fig. \ref{patchingplot}). $U=3t$, $t'=0.2t$, van Hove filling, $T=0.01t$, with discretization scheme D1 and with feedback of self-energy.} 
\label{z-delta-plot} 
\end{figure}

In general, the frequency-dependence of the imaginary part of the self-energy at these low temperatures near the instability is dominated by a linear decrease that gets steeper the closer we get to the critical scale. We can measure this angle-dependent slope $\Sigma'^\Lambda$ on the frequency window and encode it in a quasiparticle weight 
\begin{equation}
\label{zfactor}
Z^\Lambda(\mathbf{k}_F)= \left[ 1- \Sigma'^\Lambda (\mathbf{k}_F)\right]^{-1}\; ,
\end{equation}
where we define $\Sigma'^\Lambda$ as
\begin{equation}
\label{dsigma-domega}
\Sigma'^\Lambda (\mathbf{k}_F)= \frac{\Sigma^\Lambda(\mathbf{k}_F, \omega_m) - \Sigma^\Lambda(\mathbf{k}_F, -\omega_m)}{2\omega_m} \;.
\end{equation}
Here $\Sigma^\Lambda(\mathbf{k}_F, \pm \omega_m)$ are the self-energies at the Matsubara frequencies with largest absolute value $\omega_m=9\pi T$ available in discretization D1. The flow of these weights is shown in Fig. \ref{z-delta-plot}. We see that the $Z$-factor diminishes toward the instability, but in a rather weak, almost logarithmic way, such that this is possibly not the most severe effect in the self-energy. 
Besides this slope, we also notice again a step $\Delta \Sigma^\Lambda$ around zero frequency (Eq. \ref{deltasigma}), that grows rapidly toward the instability, as shown in the right plot of Fig. \ref{z-delta-plot}.
We also show curves for different locations on the Fermi surface. These different lines indicate the growth of the anisotropy toward the instability for the van Hove situation away from half filling, with stronger self-energy effects for Fermi surface points near the van Hove points. In the perfectly nested case, the anisotropy is less pronounced.

Previous works on the flow of the self-energy using different approximations have arrived at some similar but also some different conclusions regarding the self-energy flow. In Refs. \onlinecite{kataninkampf} and \cite{rohe} the anisotropy appears to be somewhat more pronounced in the sense that the Fermi surface point closest to the van Hove points showed  indications for a break-up of the quasiparticle peak into two peaks, with a spectral gap opening between them. For this to occur, one needs an additional low-energy feature in the self-energy. On the imaginary axis, this shows up as an increase of the absolute value of the self-energy imaginary part towards low frequencies, that turns around the linear slop and eventually leads to a $1/(i\omega)$-pole. In Ref. \onlinecite{kataninkampf} this could only be seen clearly by interpolating between the lowest Matsubara frequencies, an option we cannot choose in our flow with frequency-dependent coupling. However, the growing step $\Delta \Sigma^\Lambda$ can be interpreted as a precursor of this effect, as it leads to a breaking up of the simple linear decrease.  It is well possible that these subtle effects very close to the instability come out in different extent in the different approximations. Also in Ref. \onlinecite{rohe}, where the self-energy is evaluated directly on the real frequency axis, a new low-energy feature in the self-energy emerges very close to the instability and eventual splits the quasiparticle peak in two. Another aspect of this feature is a rise in the scattering rate, which is consistent with our increasing step $\Delta \Sigma^\Lambda$. On the real axis, the linear slope of of the self-energy real part goes into the $Z$-factor. In Ref. \onlinecite{rohe}, this slope after subtracting the additional low-energy feature is not strongly scale-dependent, very much like our data, where the $Z$-factor do not really dive to zero toward the instability. Since the precision of the truncated flows is not very high in this regime, we refrain from a more elaborate discussion of these differences. One message one may learn from these studies is however that the self-energy in this regime should not only be parameterized with a $Z$-factor. At least close to the instability, it develops additional low-energy structures besides this linear envelope captured by the $Z$-factor.

\section{Conclusions}
We have analyzed the low-energy properties of effective interactions and self-energies in the two-dimensional Hubbard model using a fRG approach. In contrast with previous approaches to the same model ( e.g. Refs. \onlinecite{rg1,rg2,rg3, zanchi}, and many more works cited in Ref. \onlinecite{metznerRMP}), we take into account the frequency-dependence of the interactions and the feedback of important parts of the self-energy on the flow of the latter. For this, we discretized the Matsubara frequency axis in different schemes with $N_\omega=10$ patches. For the self-energy we took into account the imaginary part (on the Matsubara axis) only, while we ignored the real part. The argument for ignoring the latter is mainly feasibility, a flowing dispersion is technically difficult to deal with, and would require additional devices to keep the particle number fixed. Our experience from previous studies \cite{UebelackerDiploma} is that the the impact of the dispersion renormalization on the instability is mainly quantitative and does not alter the regime found from the flows to strong coupling.

The main goal of this study was to see how the fRG flows of the simpler studies are altered by the straightforward inclusion of frequency-dependence and (imaginary part of the) self-energy feedback. Here, after a first sanity check on a case without flow to strong coupling, we analyzed how the critical scales for the flow to strong coupling change, and how the character of the instability in terms of the leading ordering tendency changes depending on the different approximations. The overall result is that the frequency-dependence and the imaginary part of self-energy usually decrease the critical scales by a factor at most of order one. Only near quantum critical points, where the critical scales vary strongly and become small, the change can be more drastic. Furthermore, the competition for the leading instability is also affected only quantitatively. We find the same structure of the phase diagrams and have not seen that any of the regimes found previously are wiped out by these alterations of the scheme. In particular there is a sizable regime with dominating $d$-wave pairing in the ground state.
 
Above the instability temperatures, the self-energy did not show strong signatures in its frequency- or wavevector-dependence. The basic structure could be captured by an inverse lifetime or scattering rate and a $Z$-factor for the quasiparticle weight. These parameters turned out to be weakly wavevector-dependent around the Fermi surface unless one gets close to the instability at lower temperatures. At van Hove filling and away from perfect nesting, the scattering rate above the instability comes out consistent with marginal Fermi liquid behavior, as found in previous studies\cite{lee,hlubina}. This however does not prevent the low-$T$ instabilities from occurring, i.e. even if the self-energy behaves different from a Fermi liquid, the instabilities found without inclusion of the self-energy are still present.

These findings on the behavior of the self-energy above the instability make a lot of sense, but do not reveal any unexpected physics. Certainly, the self-energy flows corroborate second-order -perturbation theory results on the non-Fermi-liquid behavior at the van Hove filling, and the low-$T$ instabilities are not changed qualitatively.  Hence the present work represents an important check for the body of knowledge has been obtained using simple fRG flows without self-energy feedback (for a review, see Ref. \onlinecite{metznerRMP}). Basically, the message we infer from the data is that frequency-dependence and self-energy effects are necessary to deal with if one is  trying to obtain truly quantitative results, but on the qualitative level, regarding the leading instabilities and orders of magnitude, these improvements do not lead to significant changes. So the simple flows without self-energy are expected to be good guides through the basic phase diagrams. Similar statements also hold for parameter trends, e.g. as function of the chemical potential of the band structure parameters. Here, we have found that these trends occur irrespective of the approximation level.

In this paper, we do not discuss explicitly the frequency structure of the effective interactions. This is quite rich and contains some interesting aspects, some of which has been discussed without self-energy feedback in Ref. \onlinecite{honphon}. Here, it is however difficult to describe the data with a few simple parameters, and less literature is available for comparison. Hence we did not elaborate on this here. It would however be interesting to compare the frequency-resolved vertex functions from this weak-coupling approach with data obtained with QMC or dynamical mean-field-like cluster methods, in order to assess the effects of strong correlations (possibly difficult for the fRG)  and larger distances (possibly less well captured by the strong-coupling approaches) on these objects.

Regarding the theoretical approach chosen here, we were able to show that it leads to useful results. On the other hand, in order to look for subtle effects near the instability such as pseudogaps or more detailed non-Fermi liquid behavior, the chosen approach appears somewhat to heavy and inflexible. The numerical effort due to a vertex that depends on three wavevectors and frequencies makes it rather cumbersome to compare different discretizations or implementations. Here, some clever reduction of the information carried along my serve useful. Our current code my serve as a guide how good these alternatives are.

At the time of writing of this paper, another work treating the self-energy feedback in fRG flows in the two-dimensional Hubbard model has appeared as a preprint\cite{giering2} which may point in the right direction. This study employs the channel decomposition of the interaction vertex proposed by Husemann and Salmhofer\cite{husemann}. The restriction to a few form factors of the interaction channels then allows the authors to reach a higher momentum space and frequency precision as in our straightforward patching scheme, and to safely go to lowest temperatures and frequencies. Furthermore they also consider the flow of the real part of the self-energy which was ignored in our scheme for simplicity. The choice of a soft frequency cutoff instead of a momentum-shell cutoff picks up ferromagnetic tendencies, which are ignored in our approach and which would be important for $t'>0.3t$ near van Hove filling. So the results of both schemes should only be compared at small $t'$. There seems to be good agreement at least in two aspects, which are the suppression of the critical scales due to self-energy effects, and the stability of the $d$-wave pairing regime with respect to inclusion of these aspects. Ref. \onlinecite{giering} also describes a non-Fermi liquid behavior of the self-energy at the van Hove-filling with exponent $3/4$ at $t'=0.355t$, which is slightly less than our second-order-like exponent near one determined for $t'=0.25t$. In comparing the pros and cons of these two related schemes, it is important to notice that our direct patching implementation without channel decomposition does allow the vertex to depend on three different wave vectors and Matsubara frequencies, while the approach in Ref. \onlinecite{giering2} only allows for certain form factors in momentum space and treats the frequency-dependence as the sum of three functions each depending on one particular frequency combinations. However, it is hoped that these constraints do not influence the results too strongly. Indeed our study shows that taking along more frequency- and momentum-structure does not change the main results and does not reveal any additional physics. Hence, it is quite likely that the channel-resolved formalism of Ref. \onlinecite{giering2} turns out to be a better compromise as it reduces the amount of information carried along in the flow in a physical meaningful way. This way one might obtain more precise data and one might be able to go on to more invalid systems without reducing the precision again. In a way, the approach chosen in this paper can be viewed as complementary, as it is less versatile but on the other hand puts less bias in the momentum and frequency-dependences of the vertex functions.

We acknowledge discussions with M. Kinza, A. Katanin, K.U. Giering, C. Husemann, M. Salmhofer, G. Klingschat and A. Eberlein. This work was supported by the DFG research units FOR 723 and 1162.

\bibliographystyle{plaindin}
\bibliography{bib}

\end{document}